# Revealing the Beauty behind the Sleeping Beauty Problem

Ioannis Mariolis

Information Technologies Institute, Centre of Research & Technology – Hellas,
6th km Xarilaou - Thermi, 57001, Thessaloniki, Greece
email: ymariolis@iti.gr

**Abstract**
A large number of essays address the **Sleeping Beauty problem**, which undermines the validity of **Bayesian inference** and Bas Van Fraassen's **'Reflection Principle'**. In this study a straightforward analysis of the problem based on **probability theory** is presented. The key difference from previous works is that apart from the **random experiment** imposed by the problem's description, a different one is also considered, in order to negate the confusion on the involved conditional probabilities. The results of the analysis indicate that no inconsistency takes place, whereas both Bayesian inference and 'Reflection Principle' are valid.

## *1. Another probability paradox?*

Probability puzzles often raise a great deal of controversy, resulting to a variety of contradictory explanations. This has also been the case for the Sleeping Beauty Problem (SBP), which was first formulated in Elga 2000.

*Elga's formulation of the problem:*
> Some researchers are going to put you to sleep. During the two days that your sleep will last, they will briefly wake you up either once or twice, depending on the toss of a fair coin (Heads: once; Tails: twice). After each waking, they will put you back to sleep with a drug that makes you forget that waking. When you are first awakened, to what degree ought you believe that the outcome of the coin toss is Heads? (Elga 2000: 143)

According to Elga's analysis, before you are put to sleep, your credence on a Heads coin toss should be 1/2, whereas when you are first awakened your credence on a Heads coin toss should be 1/3. However, no information gain (or loss) seems to take place, in order to justify the above change in your belief. Depending on whether a belief change is accepted and new evidence is assumed to be present or not, different attempts to resolve this "paradox" exist. To list a few, one may consider Elga 2000, Lewis 2001, Vaidman and Saunders 2001, Arntzenus 2002, Dorr 2002, Monton 2002, Hitchcock 2004, Weintraub 2004, Meacham 2005, White 2006, Horgan 2004, 2007, Karlander and Spectre 2010, and Hawley 2013.
    A radically different approach is presented in Groisman 2008, where it is argued that the two different degrees of belief presented in the SBP are in fact beliefs in two different propositions. Therefore, it is argued that there is no need to explain the (un)change of belief.

Finally, in Rosenthal 2009 it is identified that SBP's solution depends on conditional probabilities, which are well understood and they should be unambiguously analysable by straightforward mathematics. According to Rosenthal the difficulty in SBP seems to be that a precise mathematical interpretation of the condition involved is unclear, thus creating an obstacle to direct mathematical calculation. Rosenthal attempts to replace the problem with an equivalent one, where there is no ambiguity on the condition and then apply straightforward mathematical analysis.

In this study, a straightforward mathematical formulation of SBP is presented, employing methods of probability theory. Then, by applying direct mathematical calculations, SB's beliefs can be explicitly estimated without resulting to any ambiguities or contradictions. In contrast to Rosenthal's approach, no equivalent problem is necessary, since the involved conditional probabilities are explicitly defined in the context of corresponding random experiments. The presented analysis pinpoints the source of the controversy to a) the confusion between similar events of two different random experiments, and b) erroneously considering evidence of an event.

## 2. Defining the conducted random experiment

Modern probability theory is an axiomatic theory, where probabilities are considered in the context of a random experiment. Sleeping Beauty is participating in a Random Experiment set by the Experimenter (ERE), consisting of a single coin toss. As explained below this random experiment can also be employed to calculate the probability in question, i.e. "to what degree ought you believe that the outcome of the coin toss is Heads when you are first awakened".

Let's employ modern probability theory to describe ERE. The random part of ERE is the experimenter's coin toss. The sample space is $S = \{H, T\}$, where outcome $H$ denotes a Heads result and $T$ denotes a Tails result. Since we assume a fair coin we should assign $P(H) = P(T) = 1/2$. Assuming that you are first put to sleep on Sunday, the probability of the event "a Monday awakening occurs" is $P(Monday) = P(H) + P(T) = 1$, since SB is always awakened on Monday independent of the toss result. In fact the coin toss may actually take place after the Monday awakening. In this random experiment any knowledge that a Monday awakening occurs provides no new information since $P(Monday) = 1$. Then $P(Heads|Monday) = P(Heads,Monday) / P(Monday) = P(Heads,Monday) = P(H) = 1/2$. Moreover, $P(Monday|Tails) = P(Tuesday|Tails) = 1$.

It is clear that before you are put to sleep you can employ ERE to calculate the probability that the coin tosses Heads to $P(Heads) = P(H) = 1/2$. The ambiguous part is what happens when you are first awakened. The "when you are first awakened …" situation introduces a new kind of uncertainty, namely your current state upon awakening. However, this uncertainty is not about the ERE setup, which you still know that it is valid. Therefore, you can still consider the coin toss an ERE event and assign credence to it according to the probabilities calculated in ERE. Therefore, when you are first awakened you should still believe that the coin toss is Heads with probability $P(Heads) = P(H) = 1/2$. This result comes as no surprise, since upon your first awakening you have not gained or lost any information relevant to the coin toss. Moreover, using the same rationale, if you are informed that it is Monday you can still use ERE and calculate $P(Heads|Monday) = 1/2$, which is in agreement with Elga's proposition that if one were to learn that the waking day is a Monday, one should assign equal credence to Heads and Tails.

## *3. Defining a different random experiment*

We saw above the probabilities that can be calculated using the ERE sample space, but the source of the confusion following Elga's analysis is actually the probabilities that cannot be calculated within the ERE setup. Upon awakening you are not only uncertain about the coin toss result but also upon your state, namely whether you are awakening on a Monday or on a Tuesday. Therefore, if you ask "to what degree you ought to believe at your first awakening that it is Monday", it makes no sense to use the ERE event "a Monday awakening occurs", since it is a certain event in the ERE setup, i.e. $P(Monday) = 1$. Although, such a probability is not explicitly calculated in Elga 2000, it affects the conditional probabilities employed in that analysis.

In order to calculate probabilities of events like "to what degree you ought to believe at your first awakening that it is Monday", a different random experiment that accounts for the extra uncertainty about one's current state should be considered. Let's call this suitable random experiment SBRE and define its outcomes.

When you wake up, according to your information it could be Monday and the experimenter's coin is Heads or Tails (or not tossed yet), or it could be Tuesday and the coin was tossed Tails. Based on the available information you can model your current state as the outcome of a random experiment with two stages:
i. ERE is performed
ii. If in stage i) the coin is tossed Heads a Monday awakening is selected as your current state. If in stage i) the coin is tossed Tails either a Monday or a Tuesday awakening is randomly selected as your current state

Let Monday* denote the event "a Monday awakening is randomly selected as your current state" and Tuesday* denote the event "a Tuesday awakening is randomly selected as your current state".

Then the sample space of SBRE is S = {$H_1$, $T_1$, $T_2$}, where
$H_1$ denotes the outcome "the coin tosses Heads" and Monday*
$T_1$ denotes the outcome "the coin tosses Tails" and Monday*
$T_2$ denotes the outcome "the coin tosses Tails" and Tuesday*

Let's assign probabilities to SBRE's outcomes. $H_1$, $T_1$, and $T_2$ are mutually exclusive and jointly exhaustive, hence $P(H_1) + P(T_1) + P(T_2) = 1$. Moreover, since a Heads toss is associated only with outcome $H_1$, $P(Heads) = P(H_1)$. However, the coin toss occurs in the first stage of the random experiment, therefore its outcome does not depend on what happens on the second stage. Thus, if we assume that the coin is fair we should assign $P(Heads) = P(Tails) = 1/2$, and this implies that we should also assign $P(H_1) = 1/2$. Since $P(H_1) + P(T_1) + P(T_2) = 1$ and $P(H_1) = 1/2$, consequently $P(T_1) + P(T_2) = 1/2$. Moreover, by adopting Elga's argument (2000: 145), given a tail toss the events $T_1$ and $T_2$ are subjectively indistinguishable and should therefore be accorded the same credence. Hence, according to an indifference principle, one should assign $P(T_1|Tails) = P(T_2|Tails)$. This implies that also $P(T_1) = P(T_2)$, thus, we should assign $P(T_1) = P(T_2) = 1/4$.

Then by applying probability theory it is easy to calculate that within the SBRE setup:
$P(Monday*) = P(H_1) + P(T_1) = 3/4$

$P(Heads|Monday*) = P(Heads, Monday*) / P(Monday*) = P(H_1) / P(Monday*) = 2/3$

$P(Tails|Monday^*) = P(Tails,Monday^*) / P(Monday^*) = P(T_1) / P(Monday^*) = 1/3$

Thus, in the context of the SBRE setup, when you are first awaken, you should also believe that the probability of the coin tosses Heads, is $P(Heads) = P(H_1) = 1/2$. This is in agreement with ERE and your initial belief. Hence, both random experiments agree that after awaking credence of 1/2 should be assigned to the Heads toss.

The presented analysis demonstrates that there is no change of beliefs regarding the credence of the Heads toss before and after awakening. However, it also yields that in the context of the SBRE setup, $P(Heads|Monday^*) = 2/3$. This seems to imply that if you are informed upon awakening that it is actually Monday you should change your credence on Heads from 1/2 to 2/3. However, this is contradictory to the ERE results, which as already explained can also be employed upon awakening, yielding $P(Heads|Monday) = 1/2$ when you are informed about the day.

At first it looks like there is some disagreement on what you should believe once you learn that you awakened on a Monday, but as explained below this is not the case. The events "Monday" and "Monday*" are two different events defined within two different random experiments. Although, once you learn that it is Monday you have evidence of the event "Monday" in the ERE setup, you don't actually have evidence of the event "Monday*", i.e you don't have evidence that the SBRE event "a Monday awakening is randomly selected as your current state" has occurred. In fact, according to SBRE formulation, only 3/4 of Monday awakenings are expected to be randomly selected as your current state. Therefore, you cannot claim that you have actual evidence of "Monday*", and you cannot use the conditional probabilities $P(Heads|Monday^*)$ in SBRE to update your belief.

The above analysis reveals one of the most interesting aspects of the Sleeping Beauty Problem. Namely, it has been demonstrated that even if you are informed that it is Monday you cannot update your belief using the conditional probabilities $P(Heads|Monday^*)$. Thus, in SBP one should be careful to avoid the pitfall of performing Bayesian updating based on evidence that are not actually provided.

## 4. The "paradox's" resolution

In Elga 2000, a change of the credence assigned to the coin tossed Heads is calculated. Namely, according to Elga, before you are put to sleep the assigned credence should be 1/2, whereas it should become 1/3 upon awakening. Elga's first argument is that, if the experiment were to be repeated a large number of times, roughly 1/3 of the awakenings would be associated with Heads. Well, this is exactly the case, since Elga is referring to the ERE random experiment and counts all awakenings, even if they occur in the same trial of the experiment. However, in order to determine the probability of Heads based on awakenings, one should consider only the number of the current state awakenings that are associated with Heads. This can only be calculated in the context of the SBRE setup. If SBRE is repeated a large number of times roughly 3/4 of current state awakenings would occur on Monday and in 2/3 of them the coin toss would be Heads. Hence, 1/2 of your current state awakenings would be associated with Heads.

Elga's second argument is that:

…If (upon awakening) you were to learn that it is Monday, that would amount to your learning that you are in either $H_1$ or $T_1$. Your credence that you are in $H_1$ would then be your credence that a fair coin, soon to be tossed, will land Heads. It is irrelevant that you will be awakened on the following day if and only if the coin lands Tails — in this circumstance, your credence that the coin will land Heads ought to be 1/2. But your credence that the coin will land Heads (after learning that it is Monday) ought to be the same as the conditional credence $P(H_1| H_1$ or $T_1)$. So $P(H_1| H_1$ or $T_1) = 1/2$, and hence $P(H_1) = P(T_1)$. Combining results, we have that $P(H_1) = P(T_1) = P(T_2)$. Since these credences sum to 1, $P(H_1) = 1/3$. (Elga 2000: 145-146).

In Elga's analysis $H_1$ denotes the predicament "it is Monday and the coin tosses Heads", $T_1$ denotes the predicament "it is Monday and the coin tosses Tails" and $T_2$ denotes the predicament "it is Tuesday and the coin tosses Tails". However, in modern probability theory, predicaments cannot be used directly for calculating probabilities. Therefore, predicaments $H_1$ and $T_1$ should be associated to events of a sample space, before they can be assigned with probabilities. Within the SBRE setup predicament $H_1$ can be associated to the event $H_1$ (see Section 3) yielding $P(H_1) = P(Heads, Monday^*) = 1/2$, predicament $T_1$ can be associated to the event $T_1$ yielding $P(T_1) = P(Tails, Monday^*) = 1/4$, predicament $T_2$ can be associated to the event $T_2$ yielding $P(T_2) = P(Tails, Tuesday^*) = 1/4$, whereas "$H_1$ or $T_1$" can be associated to the event $Monday^*$ yielding $P(H_1$ or $T_1) = P(Monday^*) = 3/4$.

After learning that it is Monday you know that you are either in 'predicament' $H_1$ or 'predicament' $T_1$, but as explained in the previous section, your learning that you are awakening on a Monday provides no evidence for the occurrence of an "$H_1$ or $T_1$" SBRE event, such as Monday*. Thus, after learning that it is Monday your credence that the coin will land Heads, which is 1/2, does not have to be the same as the conditional credence $P(H_1| H_1$ or $T_1)$, which as demonstrated in Section 3 is equal to 2/3, and Elga's argument is rebutted.

However, if you are informed that the coin toss was performed on Sunday and the outcome was Tails you have evidence that a Tails result occurred and you can use SBRE to calculate the probability that today is Monday as $P(Monday^*|Tails) = P(Monday^*, Tails) / P(Tails) = P(T_1) / P(T_1$ or $T_2) = 1/2$.

The above results indicate that, although Bayesian inference is consistent, one must be very careful not to confuse:
  a) probabilities of events that belong to different random experiments
  b) facts and assumptions when deciding on the evidence of an event

Of course, once you are informed that it is Monday there is no longer need to use SBRE setup, since your current state is no longer uncertain. You can use the ERE setup, or even assume a new random experiment in which a coin is tossed only on Monday, and calculate $P(Heads) = 1/2$.

## 5. Sleeping Beauty goes to the casino

In this section the betting odds for Heads are examined. It is assumed that Sleeping Beauty is asked to bet on Heads every time she has been awakened during the ERE experiment. Although, it is not part of Elga's analysis the aforementioned betting odds are often used by thirders as a compelling argument that 1/3 is the correct answer. In particular, they argue that in order to accept the bet SB should be offered at least 2:1 odds. Namely, if the bet costs 10$ she should only accept it if the payoff is at least 30$. Coincidentally, they are correct, since in that case the expected gain is zero. What is important however, is how the expected gain is calculated. According to thirders it can be calculated by P(*Heads*)*20$+P(*Tails*)*(-10$), which yields 1/3*20$+2/3*(-10$)=0$. According to halfers it should be calculated by P(*Heads*)*20$+P(*Tails*)*(2*(-10$)), which also yields 1/2*20$+1/2*(2*(-10$))=0$. Notice that the loss is doubled in case of Tails because SB is offered the bet twice (once on Monday and once on Tuesday). If the ERE experiment is repeated a large number of times it is the second formula that provides the correct expected gain for any betting odds. Hence, if you are given 5:1 odds (e.g. if the bet costs 10$ and the payoff is 60$), thirders calculate an expected gain of 10$, whereas halfers calculate the correct expected gain of 15$. The subtle point is that every time it is Tails (hence SB loses the bet) the bet is offered twice. This is completely different than being offered a bet once with twice as much probability of loosing than winning. However, in both cases bets with 2:1 payoff result to a zero expected gain. Actually, since you are offered the same bet twice it is misleading to refer to the ratio between gain and loss amounts as odds. The actual odds for Heads is always 1:1, although due to the betting setup a 2:1 payoff is required for zero expected loss.

## *6. Conclusion*

In this study, a straightforward analysis of the SB problem, based on basic concepts of modern probability theory, has been presented. The conducted random experiment has been rigorously defined, the associated sample space has been constructed and the corresponding probabilities have been assigned. In contrast to the majority of related works, the above analysis predicts no change in the original credence, which is calculated to 1/2.

Although the conducted random experiment can be used to calculate the probability in question, Elga's analysis involved also probabilities that cannot be calculated in the context of this experiment. Therefore, a different random experiment, which accounts for the uncertainty about the day of awakening, has also been defined. The current state upon awakening can be modelled as the outcome of such random experiment, calculating the probabilities of corresponding events. The calculations about the credence assigned to the Heads coin toss upon awakening lead to the same results with those reported by the first experiment. Therefore, no change of beliefs upon awakening is predicted by the second experiment, as well. Moreover, the probabilities assigned to other events of this experiment can be used to rebut Elga's arguments on the change in the belief, demonstrating that Bayesian inference leads to no inconsistency and no paradox takes place. Moreover, in a more practical approach, a brief discussion on the betting odds for Heads has been presented. The thirder's argument that P(*Heads*)=1/3 since at least a 2:1 ratio is needed in order to accept a bet on Heads, has been rebooted.

The Sleeping Beauty problem has raised a great deal of controversy resulting to a variety of proposed explanations which are related to open philosophical questions, or even in the many worlds interpretation of quantum mechanics (Lewis 2007). However, the presented approach indicates that, as predicted by Rosenthal, the problem can be adequately addressed by modern probability theory. Moreover, the presented analysis provides insight on some of the pitfalls accompanying probability puzzles that, to the author's knowledge, have not been addressed in the published literature.

## *References*

Arntzenius, F. 2002. Reflections on Sleeping Beauty. *Analysis* 62: 53–62.
Dorr, C. 2002. Sleeping Beauty: in defence of Elga. *Analysis* 62: 292–5.
Elga, A. 2000. Self-locating belief and the Sleeping Beauty Problem. *Analysis* 60: 143–7.
Groisman, B. 2008. The End of Sleeping Beauty's Nightmare. *British Journal for the Philosophy of Science* 59: 409-16.
Hawley, P. 2013. Inertia, Optimism and Beauty . Noûs, 47: 85–103.
Hitchcock, C. 2004. Beauty and the bets. *Synthese* 139: 405–20.
Horgan, T. 2004. Sleeping Beauty awakened: new odds at the dawn of the new day. *Analysis* 64: 10–21.
Horgan, T. 2007. Synchronic Bayesian updating and the generalized Sleeping Beauty problem. *Analysis* 67: 50–9.
Karlander, K. and Spectre, L. 2010. Sleeping Beauty meets Monday. *Synthese*, 174: 397–412.
Lewis, D. 2001. Sleeping Beauty: reply to Elga. *Analysis* 61: 171–6.
Lewis, P. J. 2007. Quantum Sleeping Beauty. *Analysis* 67(293): 59–65.
Meacham, C. 2005. Sleeping Beauty and the Dynamics of De Se Beliefs. http://philsci-archive.pitt.edu/archive/00002526/ Accessed 09 April 2013.
Monton, B. 2002. Sleeping Beauty and the forgetful Bayesian. *Analysis* 62: 47–53.
Rosenthal, J.S. 2009. A mathematical analysis of the Sleeping Beauty problem. *The Mathematical Intelligencer* 31: 32-37.
Vaidman, L. and Saunders, S. 2001. On Sleeping Beauty Controversy. http://philsci-archive.pitt.edu/archive/00000324/. Accessed 09 April 2013.
van Fraassen, B.C. 1984. Belief and the will. *Journal of Philosophy 81*: 235–56.
van Fraassen, B. C. 1995. Belief and the problem of Ulysses and the Sirens. *Philosophical Studies* 77: 7-37.
Weintraub, R. 2004. Sleeping Beauty: a simple solution. *Analysis* 64: 8–10.
White, R. 2006. The generalized Sleeping Beauty problem: a challenge for thirders. *Analysis* 66: 114–9.